# Forming a highly active, homogeneously alloyed AuPt co-catalyst decoration on TiO$_2$ nanotubes directly during anodic growth

*Haidong Bian,[a] Nhat Truong Nguyen,[b] JeongEun Yoo,[b] Seyedsina Hejazi,[b] Shiva Mohajernia,[b] Julian Müller,[c] Erdmann Spiecker,[c] Hiroaki Tsuchiya,[d] Ondrej Tomanec,[e] Beatriz E. Sanabria-Arenas,[f] Radek Zboril,[e] Yang Yang Li[a],\* and Patrik Schmuki[b,e,g]\**

[a] Center of Super-Diamond and Advanced Films (COSDAF), City University of Hong Kong, Kowloon, Hong Kong, China

[b] Institute for Surface Science and Corrosion WW4-LKO, Friedrich-Alexander University of Erlangen-Nuremberg, Martensstrasse 7, D-91058 Erlangen, Germany.

[c] Institute of Micro- and Nanostructure Research & Center for Nanoanalysis and Electron Microscopy, Friedrich-Alexander University of Erlangen-Nuremberg, Cauerstraße 6, D-91058 Erlangen, Germany

[d] Division of Materials and Manufacturing Science, Graduate School of Engineering, Osaka University, 2-1 Yamada-oka, Suita, Osaka 565–0871, Japan

[e] Regional Centre of Advanced Technologies and Materials, Department of Physical Chemistry, Faculty of Science, Palacky University, 78371 Olomouc, Czech Republic

[f] Laboratorio di Corrosione dei Materiali "Pietro Pedeferri", Politecnico di Milano, Italy

[g] Chemistry Department, Faculty of Sciences, King Abdulaziz University, 80203 Jeddah, Saudi Arabia

Email: schmuki@ww.uni-erlangen.de; yangli@cityu.edu.hk

# ABSTRACT

Au and Pt do not form homogeneous bulk alloys as they are thermodynamically not miscible. However, we show that anodic $TiO_2$ nanotubes (NTs) can *in-situ* be uniformly decorated with homogeneous AuPt alloy nanoparticles (NPs) during their anodic growth. For this, a metallic Ti substrate containing low amounts of dissolved Au (0.1 at%) and Pt (0.1 at%) is used for anodizing. The matrix metal (Ti) is converted to oxide while at the oxide/metal interface direct noble metal particle formation and alloying of Au and Pt takes place; continuously these particles are then picked up by the growing nanotube wall. In our experiments the AuPt alloy NPs have an average size of 4.2 nm and, at the end of the anodic process, are regularly dispersed over the $TiO_2$ nanotubes. These alloyed AuPt particles act as excellent co-catalyst in photocatalytic $H_2$ generation - with a $H_2$ production of 12.04 μL $h^{-1}$ under solar light. This represents a strongly enhanced activity as compared with $TiO_2$ NTs decorated with monometallic particles of Au (7 μL $h^{-1}$) or Pt (9.96 μL $h^{-1}$).

# INTRODUCTION

Aligned anodic TiO$_2$ nanotube arrays have in the past decade attracted wide interest in different fields of science and technology, due to the combination of the functional features of titania with a unique 1D geometry.[1–4] The TiO$_2$ nanotube layers are typically formed by a self-organizing anodic oxidation reaction of a metallic Ti substrate in a fluoride containing electrolyte. The process allows a high control over the individual tube's geometry (length, diameter, wall thickness) and crystal structure (as-grown tubes are amorphous and can be easily crystallized to anatase or rutile by a suitable thermal treatment).[4–6]

Anatase TiO$_2$ nanotube layers have been, due to their outstanding semiconducting properties, intensively investigated for photoelectrochemical or photocatalytic H$_2$ production.[7,8] In these applications the vertical arrangement of the nanotubes on the Ti substrate provides a particularly advantageous geometry for the separation and transport of photogenerated charge carriers as well as for an optimized light management.[1,9]

However, in order to enable efficient photocatalytic hydrogen production from aqueous solutions (with or without sacrificial agents), typically for any TiO$_2$ structure the presence of a co-catalyst is needed.[10] Most effective and frequently used co-catalysts are noble metal nanoparticles, such as Au and Pt.[11,12] Particularly Pt NPs on TiO$_2$ enable a high photocatalytic activity towards H$_2$ evolution. The particles not only provide a semiconductor-metal junction that facilitates transfer for photoexcited electrons from TiO$_2$ to H$_2$O but also facilitate the hydrogen (H$_2$) formation by catalyzing the 2H$^0$ → H$_2$ recombination reaction.[10] Other noble metal NPs, such as Ag and Au, not only can work as a junction but may also aid charge transfer, *e.g.* due to surface plasmon resonance.[13–16] A particularly interesting combination are Au/Pt alloy NPs that in electrocatalytic applications recently attracted significant interest – this due to a synergistic interaction of the individual elements Au and Pt towards an enhanced reactivity, for example for methanol or carbon monoxide electro-oxidation on various electrode materials.[17–20] It is important to note that Pt and Au are essentially immiscible in the bulk (*i.e.* AuPt alloys are not thermodynamically stable)[21,22] but have been reported to form alloyed random solid solutions at size scales of <10 nm.[17,23] Findings with alloyed particles strongly differ from approaches combining the individual (elemental) noble metal catalysts on a carrier substrate. Two-metal-catalysts on a support (*e.g.* Au plus Pt on $TiO_2$ powders, tubes, rods, NPs) can combine the individual plasmonic effects of Au with high catalytic effects of Pt.[24,25] In contrast, in cases where true alloying has been reported,[26,27] mainly electronic effects of AuPt, *i.e.* alteration in the work function of the alloy (as opposed to Au or Pt), are considered to enhance $H_2$ formation. In these works, alloyed AuPt particles are produced by exploring $TiO_2$ powders or tubes to a chemical reduction or impregnation approach from noble metal containing solutions.

In the present work we show for the first time the synthesis of homogeneously alloyed AuPt NPs during an anodization process of a suitable Ti-X solid solution substrate (X = Au, Pt). The NPs are directly picked up by the growing nanotubes (*i.e.* in a one-step approach), leading to a uniform AuPt-NP self-decoration. If investigated for photocatalytic $H_2$ production, these AuPt alloy NP decorated $TiO_2$ NTs exhibit a significantly higher performance than monometallic Au or Pt NPs decorated on $TiO_2$ NTs.

## **RESULTS AND DISCUSSION**

For our experiment, we use a substrate that contains a sufficiently low noble metal content (0.1 at% of Au and 0.1 at% of Pt) to provide a precipitation-free solid noble-metal solution. During anodizing, the Ti matrix is oxidized (dissolved and converted to $TiO_2$ NTs)[4] while Au and Pt combine and aggregate to NPs – we will show below how at the metal/oxide interface homogeneously alloyed NPs are continuously formed (as illustrated in Figure 1a) and then are integrated in/on the growing oxide tubes. Continuous nanotube growth results in a uniform decoration over the surface of the $TiO_2$ nanotube walls.

Figure 1b and c show top and side view scanning electron microscopy (SEM) images of AuPt alloy NP decorated $TiO_2$ nanotube arrays that were formed after anodizing the TiAuPt substrate-alloy in 0.2 M HF/1 M $H_2O$ in ethylene glycol solution at 120 V for 105 min. It is evident that the AuPt NPs are present on the surface and on the walls of the NTs, and also in the bottom region of the tubes (Figure S1-S3). A uniform distribution of the AuPt alloy NPs is also evident from the high angle annular dark field - scanning transmission electron microscopy (HAADF-STEM) image in Figure 1d and HR-TEM images in the inset of Figure 1d and in Figure S4. TEM image analysis shows an average diameter of the loaded NPs to be 4.2 nm ± 1 nm, as evaluated in Figure S5. Most importantly, the elemental mapping results (transmission electron microscopy - energy dispersive spectroscopy TEM-EDS) of nanoparticles (Figure 1f-h) as well as a line scan of a single alloy particle (Figure 1i) show that Au and Pt are homogeneously distributed across the entire particle (Figure 1e-i), demonstrating that indeed the single particle consists of a homogeneous alloy. Additional TEM-EDS (Figure S6) taken at the top, middle and bottom of the tube shows that nanoparticles are present in an alloyed form over the entire tube length. Quantification of the TEM-EDS analysis gives an Au to Pt atomic ratio of approximately 1:1 (see Table S1), which is well in line with the original stoichiometric ratio of Au and Pt in the TiAuPt alloy substrate. The XPS depth profile (Figure 1j) of the outermost part of the tubes and a glow discharge optical emission spectroscopy (GDOES) profile over the entire tube length (Figure 1k) give an overview of the particle distribution over the tube layer. From Figure 1j, it is apparent that Au and Pt (*i.e.* the amount of the alloyed particles) show a slight enrichment in the outermost part of the sample.

Except for this approximate 50 nm thick enriched layer, the Pt and Au concentrations level off to 0.6 at% and 0.5 at%, respectively, over the entire remaining tube length (Figure 1k). The enrichment of Pt and Au in the outermost part of the tubes can be ascribed to a loss of $TiO_2$ due to mild etching of the tube walls in the fluoride electrolyte.[28]

Figure 2a illustrates the formation process of these alloyed particles. In the metallic substrate, at the selected low concentration (Au-0.1 at% and Pt-0.1 at.%), the noble metals are fully soluble in Ti in accord with literature.[29] In our case this is evident from XRD analysis of the substrate (Figure 2g) combined with SEM image of the TiAuPt substrate after attentive fine-polishing (Figure 2e). Clearly the surface is homogeneous and no precipitation of Au, Pt, or other phases is observed in SEM. The XRD for the TiAuPt alloy is entirely featureless in the Au and Pt region – this also holds for other substrates (TiAu and TiPt) that were used for reference (Figure 2g). Therefore, the substrates do not contain precipitates of noble metals (but the elements are fully soluble). The observed noble metal particles after anodizing must be formed during the anodization process and then decorate the oxide wall (see Figure S2 for reference TiAu and TiPt substrate).

Figure 2b-d provide detailed SEM images taken at the alloy/nanotube interface. It can be observed that NPs indeed are formed at the anodizing front, *i.e.* where the Ti-substrate metal is converted to oxide. At this interface the liberated noble metal atoms agglomerate to an alloy particle. These alloyed particles then become picked up by the growing tube and remain as a wall decoration. The metal/oxide interface and the initial alloy formation process was studied in more detail using highly polished surfaces as in Figure 2e that are exposed to a chemical oxidation and etching treatment (see experimental section for details). Upon oxidation (Ti-dissolution) one obtains NPs that are regularly distributed over the surface of the TiAuPt alloy (Figure 2f), *i.e.* NPs form and aggregate on the metal substrate (see also Figure S7).

In the Ti-X alloy (X = noble metal), the noble metal is dissolved; therefore for these alloys (before any treatment) no noble metal peak in XRD can be detected (Figure 2g). If the metal surface is exposed to an oxidation – oxide-dissolution treatment with increasing treatment time, noble metal particles start forming that steadily increase in density. For these surfaces, XRD (Figure 2h) clearly shows for a Ti-Au substrate the formation of Au particles, for a Ti-Pt substrate the formation of Pt particles, and for Ti-Au-Pt the formation of alloyed AuPt particles. This first dealloying step (selective dissolution) takes place where the non-noble metallic component is selectively lost. Noble metal nanoparticles aggregate due to surface diffusion.[30–32] This model experiment illustrates that alloy particles can indeed easily form at the dissolving metal/oxide interface. In the full anodizing process, due to the low solubility of noble metals in $TiO_2$, the metal cannot be integrated as dissolved species in the oxide and noble metal aggregates cannot grow to large clusters as they are continuously removed from the surface by the growing oxide nanotubes. The final nanoparticle decoration is uniform with 4.2 nm sized particles spaced by 7.0 nm. From X-ray photoelectron spectroscopy (XPS) data of Figure 1j and EDS data for the entire tube layer, the alloy shows an approximately equal atomic concentration of Pt and Au (0.49:0.47 at%, Figure 3a). These data show an overall enrichment of Au/Pt in the tubes compared to the substrate, which can be ascribed to loss of titania to the electrolyte during anodizing.[33]

To investigate the photocatalytic activity of the alloyed NPs as a co-catalyst on $TiO_2$ NTs towards $H_2$ evolution and to compare this to the individual elements, we produced also noble metal decorated $TiO_2$ NTs with the same geometry and comparable particle decoration (see Figure S2) using single metal alloys, that is TiAu (0.2 at% Au) and TiPt (0.2 at% Pt); details of TiAu and TiPt anodization are given in the Materials and Methods and Figure S3. Figure 3b shows the $H_2$ evolution of the AuPt alloy decorated NTs, compared with Au and Pt NPs decorated $TiO_2$ NTs under open circuit conditions using UV light (325 nm, HeCd laser, 60 mW cm$^{-2}$). It is evident that the AuPt alloy NP decorated TiO$_2$ NTs show clearly the most effective H$_2$ evolution with a rate of 255 µL h$^{-1}$, while TiPt and TiAu NTs exhibit a H$_2$ production rate of 185 µL h$^{-1}$ and 81 µL h$^{-1}$, respectively (while plain tubes provide only 0.64 µL h$^{-1}$). Furthermore, repeated photocatalytic experiments (Figure 3c) show that the activity of the AuPt alloy NP decorated TiO$_2$ NTs is stable, *i.e.* over the entire investigated time, the H$_2$ production rate is steady and no co-catalyst flake off or deterioration could be observed.

In order to evaluate the photocatalytic activity towards H$_2$ evolution under solar light, the nanotube samples were irradiated under AM 1.5 illumination using a calibrated solar simulator with an intensity of 100 mW cm$^{-2}$. Figure 3d exhibits the H$_2$ production amounts of TiAu, TiPt and TiAuPt NTs after 5 h of illumination. Also under these illumination conditions, it is evident that the most effective H$_2$ production is observed from the AuPt alloy NP decorated TiO$_2$ NTs with 12.0 µL h$^{-1}$ while TiAu and TiPt NTs exhibit a H$_2$ production of 7.0 and 9.96 µL h$^{-1}$, respectively. In literature, some work has demonstrated that a combined loading of individual Au and Pt particles on TiO$_2$ (note: non-alloyed) can lead to a higher photocatalytic or electrocatalytic activity than using only one catalyst species.[24,34] This combined effect of individual NP of Pt and Au for visible light illumination has several times been ascribed to contribution of plasmon resonance of Au, which in the visible range results to an improvement in the photocatalytic performance.[24,27,35] In our case, however, the alloyed AuPt particles do not yield a plasmon band in the visible range (see Figure S8 and S9); only for the pure Au decorated tubes a plasmon related feature centered at 550 nm can be observed.

Therefore the high efficiency of AuPt alloy NP decorated TiO$_2$ NTs under both illumination types must be ascribed to non-plasmon related synergistic effect between the Au and Pt in the alloyed NPs. The main electronic and chemical features altered in an alloy are: *1)* shift in the $H_2$ → $H^0$ equilibrium, *i.e.* affecting hydrogen recombination and dissociation kinetics of the catalyst[36,37] and *2)* a different work function of the alloy and thus a change in the Schottky junction of the metal/oxide interface.[26] XPS data show in the low energy XPS region (Figure S9) that noble metal decoration on the $TiO_2$ NTs induces additional electronic states in the bandgap. Clearly, the population of the states (detected electrons) from the TiAuPt NTs is higher than that of the TiAu and TiPt NTs (Figure S11).[26] This is in line with several studies that find a charge transfer between the alloying elements due to the difference in the electronegativity between Pt and Au.[23,38,39] This affects the electron density distribution, the d-band filling of Pt, and the overall Fermi level of the NP.[26,40] This, in turn, is well in line with the XPS data in Figure S8 that indicates the highest population of intra gap states caused by alloyed particles. We thus consider that in line with various literature reports[24,26,36] these electronic effects are responsible for the observed enhanced co-catalytic activity for photocatalytic $H_2$ evolution from $TiO_2$ NTs. Additionally it should be noted that the AuPt NPs are remarkably thermally stable, *i.e.* even thermal annealing process used in this work (450°C, 1 h) to crystallize amorphous material to anatase did not lead to a de-mixing of the alloyed particles, this supports concepts that ascribe the stabilization of alloyed NP (<10 nm) to a thermodynamic rather than a kinetic effect.[23]

## CONCLUSIONS

The present work shows that homogeneous AuPt alloy NPs can be formed by anodizing low concentration noble-metal Ti alloys. If this approach is used during anodic formation of $TiO_2$ NTs, then these alloy particles are formed *in situ* and loaded on $TiO_2$ NTs. Formation and alloying of the AuPt NPs occurs at the oxide-metal interface. In our case, alloy particles have an average size of ~4.2 nm and are regularly distributed over the entire growing NTs. The AuPt alloy NP decorated $TiO_2$ NTs exhibit a significantly higher photocatalytic $H_2$ evolution than pure Au or Pt decorated $TiO_2$ under UV light irradiation (1.5 times higher than of Pt decorated $TiO_2$ NTs and more than 3 times higher than Au decorated $TiO_2$ NTs). The approach introduced here that provides a one-step anodic alloy nanoparticle synthesis (of a noble metal combination that in the bulk is not miscible) is promising not only for $TiO_2$ NT decoration and photocatalytic applications, but can likely be transferred to analogous systems, to create noble metal alloy decorated structures from bulk-immiscible noble particle combinations.

## MATERIALS AND METHODS

*Growth of $TiO_2$ nanotubes*: Ti alloy sheets (0.2 mm, Hauner Metallische Werkstoffe, Germany) with 0.1 at% Au and 0.1 at% Pt were used for anodization. Before anodization, the foils were sonicated in acetone, ethanol and deionized water, followed by drying with nitrogen gas. The self-ordering anodization was carried out in a home-made two-electrode cell in ethylene glycol (EG) solution containing 0.2 M HF / 1 M $H_2O$ at 120 V for 105 min with a LAB/SM1300 as a power supply. The thickness of $TiO_2$ nanotube layers was ~7 μm. After the anodization process, the samples were washed in ethanol, dried in a nitrogen stream, and annealed at 450 ºC for 1 h in air to crystallize the materials. For comparison, TiAu (0.2 at% Au) and TiPt (0.2 at% Pt) alloy substrates were also anodized in EG solution containing 0.2 M HF / 1 M $H_2O$ at 120 V for 105 min (the voltage was ramped at 120 V $s^{-1}$).

*Characterization of the Structures*: Field-emission scanning electron microscope FE-SEM images were acquired with a Hitachi SEM FE 4800. Energy-dispersive X-ray spectroscopy (EDAX Genesis, fitted to SEM chamber) was used for the chemical analysis. EDS measurements were conducted at 20 keV. The composition and chemical states were characterized by X-ray photoelectron spectroscopy (XPS, PHI 5600, US) and peak positions were calibrated on the Ti2p peak at 458 eV. For valence band measurements a monochromated Al Kα X-ray radiation source at 300 W with resolution of 0.1 eV was used. UV-Vis diffuse reflectance spectra (DRS) were recorded on a LAMBDA 950 UV-vis spectrophotometer (Perkin Elmer, Beaconsfield, UK) with an integrating sphere ($BaSO_4$ standard white board was used as reference). Transmission electron microscopy was carried out at a double-corrected FEI Titan$^3$ Themis equipped with a SuperX detector. X-ray diffraction (XRD), performed with an X'pert Philips MPD (equipped with a Panalytical X'celerator detector) using graphite monochromatized Cu $K_α$ radiation (λ = 1.54056 Å), was used to analyze the crystallographic properties of the materials. Elemental depth profile analysis was performed using a Horiba Jobin-Yvon 5000 RF glow discharge optical emission spectroscopy (GDOES) instrument in an argon atmosphere of 650 Pa by applying an RF of 3000 MHz and a power of 27 W. Light emissions of characteristic wavelengths were monitored throughout the analysis with a sampling time of 0.1 s to obtain depth profiles. The signals were detected from a circular area of approximately 4 mm diameter.

*Chemical oxidation*: for model substrates, the TiAuPt foil was mechanically grounded down to 320 grit abrasive SiC paper. Then rough polishing was done with 9μm METADI diamond on an ULTRA-PAD™ cloth. In this step METADI Fluid was used instead of water for lubrication. Lastly, samples were final polished with MASTERMET 2 colloidal silica on a MICROCLOTH pad for 30 minutes. The rotation speed for all the steps were kept at 120 rpm. Kroll's solution (3 mL HF, 6 mL $HNO_3$, and 100 mL water) was prepared and then the sample immersed in solution for 10, 30, 60 and 180 s, afterwards washed in deionized water and dried with hot air.

*Photocatalytic Experiments*: For photocatalytic $H_2$ production, the annealed samples were immersed in a quartz tube containing water/ethanol (80 vol/20 vol) that was previously purged with $N_2$ for 15 min. After that, a 200 mW HeCd laser ($I_{light}$ = 60 mW cm$^{-2}$, λ = 325 nm, Kimmon, Japan) used as UV light and an AM 1.5 solar simulator (300 W Xe, Solarlight, 100 mW cm$^{-2}$) used as visible light were irradiated on the samples. In order to measure $H_2$ production, gas samples were analyzed by gas chromatography (GCMS-QO2010SE, SHIMADZU). The GC was equipped with a thermal conductivity detector (TCD), a Restek micropacked Shin Carbon ST column (2 m × 0.53 mm). GC measurements were carried out at a temperature of the oven of 45 °C (isothermal conditions), with the temperature of the injector set up at 280 °C and that of the TCD fi xed at 260 °C. The flow rate of the carrier gas, *i.e.* argon, was 14.3 mL min$^{-1}$.

## Associated content

**Supporting Information**. Additional SEM and TEM images, TEM-EDS result, size distribution of the nanoparticles, XPS spectra and diffuse reflectance spectra.

The authors declare no competing financial interest.

## Acknowledgements

The authors would like to acknowledge the ERC, the DFG and the DFG cluster of excellence, project EXC 315, the DFG funCOS and the Operational Programme Research, Development and Education - European Regional Development Fund, project no. CZ.02.1.01/0.0/0.0/15_003/0000416 of the Ministry of Education, Youth and Sports of the Czech Republic for financial support.

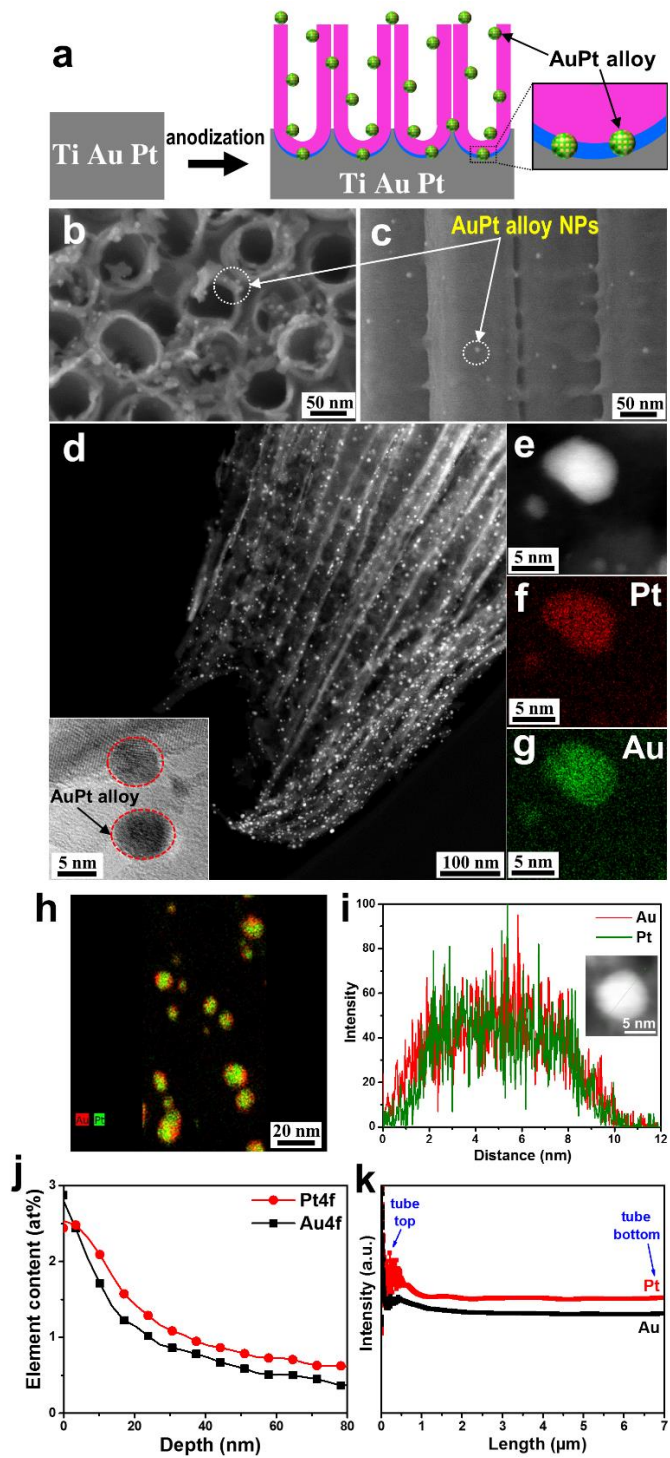

**Figure 1.** (a) Scheme of alloy anodization. (b) Surface and (c) cross-sectional SEM images. (d) Low- and (e) high-magnification HAADF-STEM images of AuPt-decorated $TiO_2$ nanotube arrays of Ti-Au(0.1 at%)-Pt(0.1 at%) anodizing in 0.2 M HF/ 1 M $H_2O$ ethylene glycol electrolyte. Inset in (d) shows the HR-TEM image of AuPt nanoparticle on $TiO_2$ NTs. (f) and (g) elemental mapping result of one AuPt alloy nanoparticle on the $TiO_2$ surface. (h) Elemental mapping (Au and Pt) of several AuPt alloy nanoparticles on $TiO_2$ nanotubes. (i) EDS line scan of single AuPt alloy particle. (j) XPS depth profile and (k) GDOES of Au and Pt on the AuPt-decorated $TiO_2$ NTs.

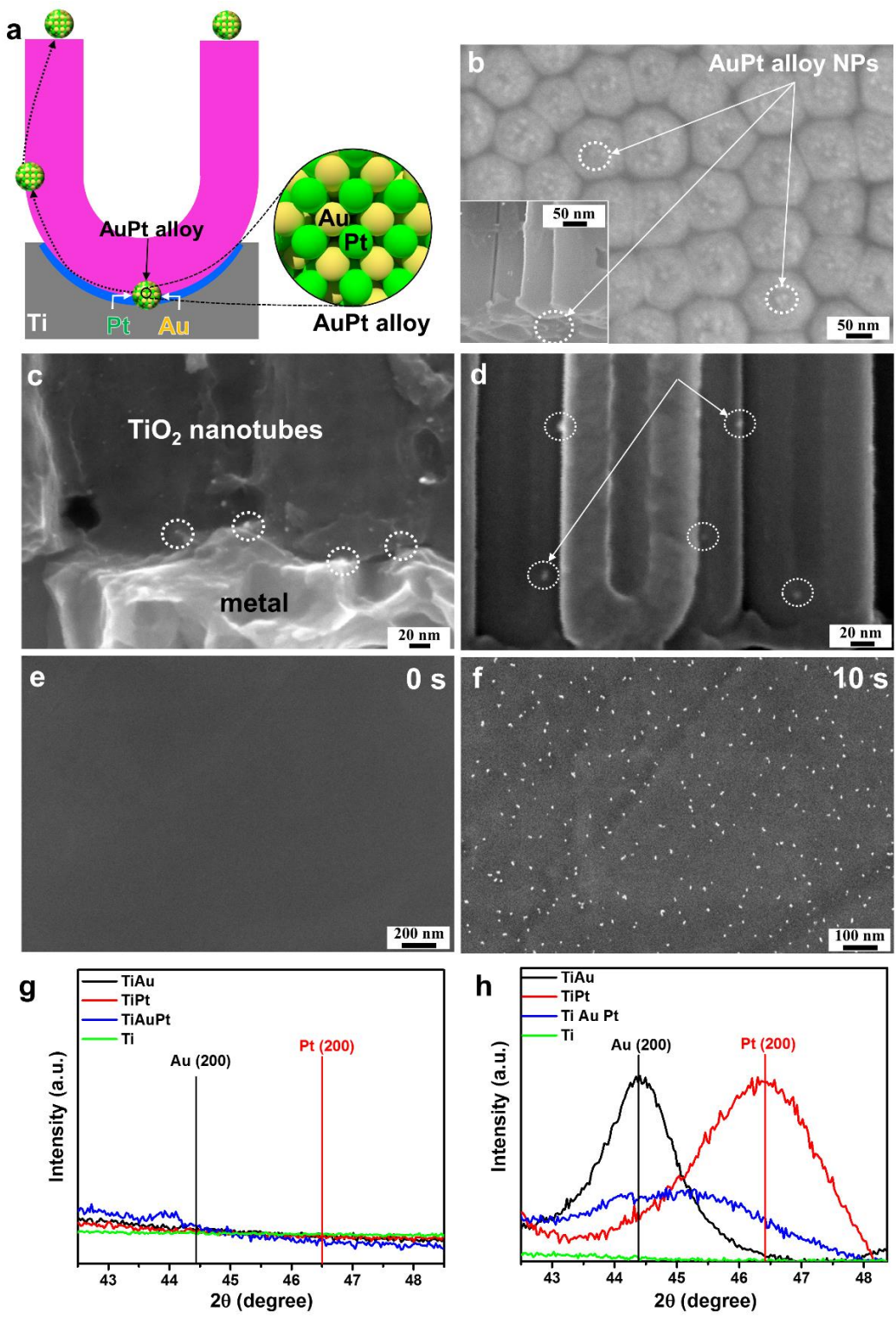

**Figure 2.** (a) Scheme of the formation of the AuPt alloyed NPs. SEM images of: (b) bottom view and (c), (d) cross-section view of TiO$_2$ NTs decorated with AuPt alloy NPs. The inset shows the cross-section view of (b). SEM images of (e) TiAuPt foil after polishing and (f) TiAuPt foil after chemical oxidation. (g) XRD patterns of TiAu, TiPt and TiAuPt substrates before chemical oxidation. (h) XRD patterns of Ti, TiAu, TiPt and TiAuPt substrates after chemical oxidation for 10 s.

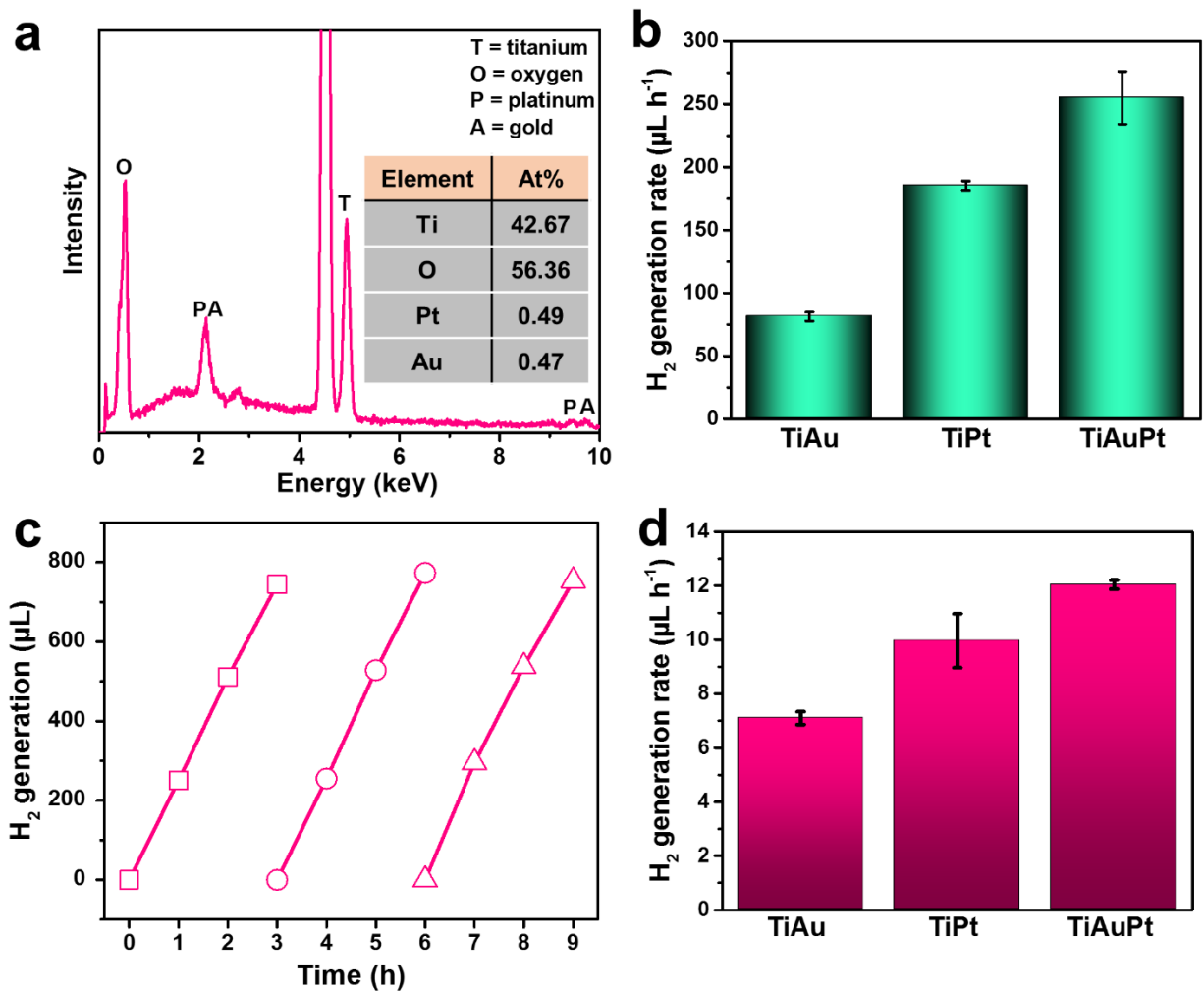

**Figure 3.** (a) EDS spectrum and elemental concentration (inset) of TiAuPt NTs. (b) Photocatalytic $H_2$ production of NP-decorated $TiO_2$ NTs formed by anodizing TiAuPt, TiAu and TiPt alloys under UV light ($\lambda$ = 325 nm). (c) Stability of photocatalytic $H_2$ production of TiAuPt NTs under UV light. (d) Photocatalytic $H_2$ production of TiAuPt, TiAu and TiPt NTs under solar simulator (AM 1.5G, 100 mW cm$^{-2}$).

**TOC**

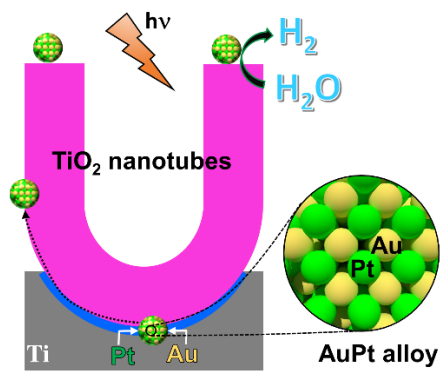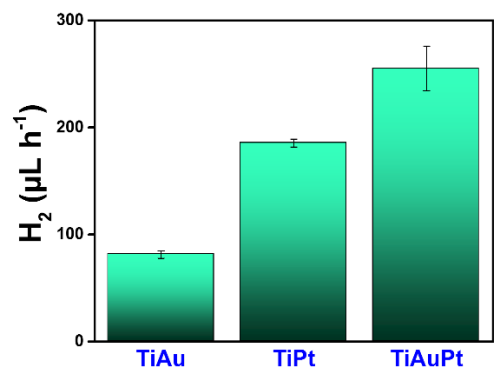